\documentclass[twocolumn, amsmath, superscriptaddress, amsfonts,prb]{revtex4-2}
\usepackage{graphicx}
\usepackage{url}
\usepackage{appendix}
\pdfoutput=1

\begin{document}

\title{Barometric equation and extent of planetary atmospheres}

\author{Marcos Grimsditch}\email{mgrimsditch@me.com}\affiliation{Independent Researcher, Port Angeles, WA 98362, USA}
\author{Victor G. Karpov}\email{victor.karpov@utoledo.edu}\affiliation{Department of Physics and Astronomy, University of Toledo, Toledo,OH 43606, USA}

\begin{abstract}
The barometric equation predicts the molecular concentration $n(z)$ exponentially decaying with altitude $z$.
Because the mean free path $l=1/n\sigma$ increases exponentially, at high altitudes $z$, the equation is no longer within the domain of applicability of the standard kinetic theory. \cite{grimsditch} Here, we predict the dependence $n(z)\propto z^{-2}$ for the case $l\gg L$ in uniform gravity and $n(z)\propto [\ln(1+z/R_0)]^{-2}$ when $z\gg R_0$ for a planet of radius $R_0$. It corresponds to a non-stationary planetary atmosphere with hydrogen accretion. The accretion is accompanied by a release of gravitational potential energy that could be the elusive source driving the formation of stellar coronas.  Other consequences include slowly decaying tails of planetary atmospheres and periodical hydrogen explosions of white dwarfs.
\end{abstract}

\maketitle

\section{Introduction. Barometric equation}\label{sec:intro}
The problem we treat here is the behavior of a gas when gravitational energies become comparable to thermal energies, which situation exists in upper planetary atmospheres due to the exponentially increasing mean free paths.  We show that the atmosphere density in that region is inversely proportional to the square of the altitude and further the square of the logarithm of it, which explains the observed long tails of the atmosphere density. Our analysis predicts a non-stationary atmosphere with hydrogen accretion accompanied by a release of gravitational energy contributing to stellar and planetary coronas.

The barometric equation,
\begin{equation}\label{eq:bareq}n(z)=n_0\exp\left(-\frac{z}{L}\right)\quad {\rm with}\quad L=\frac{k_BT}{mg}\end{equation} follows from the Boltzmann-Gibbs distribution $n=n_0\exp[-U(z)/k_BT]$ with the potential energy $U=mgz$ and can be expressed in terms of pressure $P=nk_BT$. While this approach does not invoke the mean free path $l$ or other kinetic concepts, it is limited to the condition of gas theory, $l\ll L$.

The latter limitation becomes more explicit when we present a derivation based on the kinetic concepts based on the stationary continuity equation,
\begin{equation}\label{eq:cont}-D\frac{dn}{dz}+vn=0.\end{equation}
For the diffusion coefficient, we use the standard approximation,
\begin{equation}\label{eq:difco}D=v_Tl=v_T/(n\sigma ) \quad {\rm with}\quad v_T=\sqrt{k_BT/m},\end{equation}
where $v_T$ is the thermal velocity and $\sigma$ is the cross section of molecular interactions. The drift velocity $v=\mu F$ with $\mu$ being the mobility and $F=-mg$ force according to the Einstein relation,
$\mu = D/k_BT$. Eq. (\ref{eq:cont}) reduces then to the differential equation $Ddn/dz+(D/k_bT)mgn=0$ yielding the barometric formula.

It may be worth noting here that in the long range case $R\gg R_0$ the force $F=mg$ must be replaced with the universal gravity,
\begin{equation}F=mg\frac{R_0^2}{R^2}\end{equation}
where $R_0$ and $R$ are respectively the planet radius and distance to its center. The latter modification is easily accounted for leading to the correspondingly modified barometric formula, \cite{chamberlain1978}
\begin{equation} \label{eq:bareq2}n=n_0\exp\left[-\frac{R_0}{L}\frac{(R-R_0)}{R}\right].\end{equation}
This long range case will be further discussed in Sec. \ref{sec:LRB} and Appendix \ref{sec:boltz} below.

As expressed through gas parameters, the criterion  $l\ll L$ becomes
\begin{equation}\label{eq:criterion}\frac{k_BT}{mgl}\gg 1.\end{equation}
That criterion can be interpreted as a limitation on gas concentration $n$ that must be high enough for the barometric equation to apply. Introducing zero altitude mean free path $l_0=1/n_0\sigma$, an approximate formula for the upper bound domain for, barometric formula region is given by,
\begin{equation}\label{eq:zcrit}z=z_c\approx L\ln \left(\frac{L}{l_0}\right).\end{equation}
Its corresponding concentration is roughly estimated as,
\begin{equation}\label{eq:nc}n_c=n_0\exp(-z_c/L)\approx n_0(l_0/L).\end{equation}

To keep a link to a specific case we consider Earth's hydrogen atmosphere, for which $n_0\sim 10^{14}$ cm$^{-3}$, $\sigma\sim 10^{-16}$ cm$^{2}$, $g=10^3$ cm/s$^2$, $m\approx 3\times 10^{-24}$ g. These yield the parameters given in Table \ref{tab:par}.
\begin{table}[h]
\caption{Some parameters of the Earth hydrogen atmosphere .}
\begin{tabular}{|c|c|c|c|c|c|}
  \hline
  parameter  &$l_0$& $L$ &$z_c$ &$n_0$ &$n_c$ \\\hline
  estimate & 1 m &  120 km &1500 km & $10^{14}$ cm$^{-3}$& $10^{7}$ cm$^{-3}$\\ \hline

  \hline
\end{tabular}\label{tab:par}
\end{table}

The domain of lower concentrations at $z>z_c$ belongs in the rarified systems.
In what follows, we modify our derivation for the case of rarified gases at $z>z_c$ where the criterion in Eq. (\ref{eq:criterion}) fails.

\section{Instability in barometric distribution}\label{sec:inst}
Here, we present an evidence of an instability in the Barometric equation. We proceed from the kinetic Boltzmann equation for a one-component gas singlet distribution function $f_j^{(1)}({\bf r},{\bf v}_j,t)$, omitting index $j$ and assuming the hydrogen gas as the lightest,
\begin{equation}\label{eq:Boltz}
\frac{\partial f^{(1)}}{\partial t}+v\frac{\partial f^{(1)}}{\partial z}-g\frac{\partial f^{(1)}}{\partial v}=I,\end{equation}
$\int dzdvf^{(1)}$ gives the total number of particles. In the relaxation time approximation, the collision integral $I$ is represented as
\begin{equation}\label{relax}I=-\frac{\delta f}{\tau},\quad {\rm with} \quad \delta f=f^{(1)}-f_0^{(1)}\end{equation}
where
\begin{equation}\label{eq:f0}f_0^{(1)}=const\times \exp\left(-\frac{mv^2}{2k_BT}-\frac{mgz}{k_BT}\right)\end{equation}
is the equilibrium distribution function satisfying Eq. (\ref{eq:Boltz}) with $\partial f/\partial t =0$, $T=const$ and $I=0$. \cite{klim,be,struchtrup}

We seek instabilities of the distribution function, in the form,
\begin{equation}\label{eq:ansatz}\delta f=\tilde{f}(t,z)\exp (-mv^2/2k_BT),\end{equation}
which reduces Eq. (\ref{eq:Boltz}) to the following.
\begin{equation}\label{eq:Boltz1}
\frac{\partial \tilde{f}}{\partial t}+v\frac{\partial \tilde{f}}{\partial z}-g\frac{mv}{T}\tilde{f}=-\frac{\tilde{f}}{\tau}\end{equation}

Following the standard stability analysis we look for its solution as a Fourier expansion,
\begin{equation}\label{eq:Fourier}\tilde{f}=\sum _{\omega , q}a_{\omega ,q}^{(0)}\exp(-i\omega t-iqz)\end{equation}
where $\omega$ and $q$ are the frequency and wave number of a  partial wave with the initial ($t=0$) amplitude $a_{\omega ,q}^{(0)}$. Substituting Eq. (\ref{eq:Fourier}) into Eq. (\ref{eq:Boltz1}) can result in some partial waves having the positive real parts of $\-i\omega$, i. e. $\Re {(-i\omega )}\equiv \lambda >0$; such a wave is indicative of temporal instability of such waves.

Substituting the ansatz of Eq. (\ref{eq:ansatz}) into Eq. (\ref{eq:Boltz1}) yields the dispersion law,
\begin{equation}-i\omega =\frac{mgv}{k_BT}-\frac{1}{\tau} +ivk \quad {\rm i. e.} \quad \lambda =\frac{mgv}{k_B T}-\frac{1}{\tau}.\end{equation}
The instability criterion $\lambda >0$ takes the form
\begin{equation}\label{eq:M}\frac{mgl}{k_BT} >1\end{equation} where $l=v\tau$ is the mean-free-path. It coincides with our former result in Eq. (\ref{eq:criterion}).
The complementary part of spectrum with $\lambda <0$ represents decaying fluctuations  of no interest here. Note that the above instability is related to the term $\partial f^{(1)}/\partial v$ describing the evolution of velocity distribution (across the atmosphere) accounted for by the Boltzmann equation (\ref{eq:Boltz1}).

One additional observation is that the system remains stable for low altitudes ($l$ small enough to keep $\lambda <0$), however it becomes progressively unstable with $z\gg z_c$. This corresponds to the downward stream of molecules towards the edge of stability $z_c$.

\section{Modification for rarified gases}\label{sec:rar}
In a dilute gas at $z>>z_c$, most of the molecules are not interacting with any other molecule and are just travelling along between collisions. Because of this, the macroscopic behavior of a gas depends only upon a singlet distribution function $f_j^{(1)}({\bf r},{\bf v}_j,t)$ where the subscript $j$ denotes the singlet distribution function of species $j$. One could, if helpful, treat a gas as pure H$_2$. At about $z\approx 100$ km the numerical density of H$_2$ is approximately equal to the density of N$_2$.  At $z\approx 200$ km the H$_2$ density is already $10^5$ higher than that of N$_2$.

Following Ref. \cite{mcquarrie} (p. 406, )we introduce the mass average (stream) velocity ${\bf v}_0$.
\begin{equation}\label{eq:aver}{\bf v}_0({\bf r},t)=\frac{\sum _jm_jn_j{\bf v}_j}{\sum _jm_jn_j}\end{equation}
The momentum flux of the gas is the same as if all the molecules were moving with velocity ${\bf v}_0$.
Simultaneously, we define the velocity ${\bf V}_j$ of a molecule relative to the stream velocity,
${\bf V}_j={\bf v}_j-{\bf v}_0. $
The average of this peculiar velocity is the diffusion velocity,
\begin{equation}\label{eq:avepec}{\overline{\bf V}}_j=\frac{1}{n_i}\int ({\bf v}_j-{\bf v}_0)f_j^{(1)}({\bf r},{\bf p}_j,t)d{\bf v}_j.\end{equation}
Its average over all species is zero, $\sum _jm_jn_j{\overline{\bf V}}_j=0$.
Due to its stochasticity, the characteristic rms value of ${\overline{\bf V}}_j$ can be identified as the thermal velocity,
\begin{equation} \label{eq:thermvel}  \sqrt{\langle({\overline{\bf V}}_j)^2\rangle}=v_T=\sqrt{k_BT/m}\end{equation}
where $T$ is treated as the external parameter. Therefore, for the diffusion coefficient, we use the standard approximation,
\begin{equation}\label{eq:difco1}D=v_T/(n\sigma )\equiv v_Tl.\end{equation}

The stream velocity $v$ is dominated by the gravity and its average value is given by
\begin{equation}\label{eq:strvel}v_0=v_g=\sqrt{gl/2}.\end{equation}
Eq. (\ref{eq:strvel}) tacitly assumes zero constant velocity $v_0$ in the general expression for accelerated motion, $v=v_0+\sqrt{gl/2}$, hence, acceleration starting on average with zero initial velocity after each collision. According to our classification, the random collision-related contributions must be assigned to peculiar velocities; hence, $v_0\equiv \langle v_0\rangle =  0$, intuitively obvious because the average momentum change is zero in every pair collision.

While Eq. (\ref{eq:strvel}) appears obvious from the point of view of the classical mechanics, its meaning may raise questions related to statistical mechanics, particularly about the relation between $v_g$ and the mobility ($\mu$ ) related drift velocity $v_\mu =\mu mg$. It can be shown (see Appendix \ref{sec:streamvel}) that Eq. (\ref{eq:criterion}) enables the standard mobility concept with velocity $v_\mu (\gg v_g)$, while $v_g (\gg v_\mu )$ must be used in the opposite limit $mgl/k_BT\gg 1$ describing the rarified gas.

Noting that $p_g=mv_g$ is the stream component of the momentum of one molecule, it is straightforward to see that its related momentum flux
\begin{equation}\label{eq:momflux}{\cal P} =(nv_g)p=mnv_g^2\end{equation} remains constant as a function of coordinate because $v_g\propto 1/\sqrt{n}$.
The corresponding energy flux $\Phi =(nv_g)(mv_g^2/2)=(mg/4\sigma)\sqrt{g/2n\sigma}$. Substituting here $n=n_c$ from Eq. (\ref{eq:nc}) yields an estimate for the total power per area brought by the accretion,
\begin{equation}\label{eq:power}w=\frac{mg}{4\sigma}\sqrt{\frac{gL}{2}}.\end{equation} Substituting here the above mentioned Earth parameters, yields a numerical estimate, $w\sim 0.1$ $\mu$W/cm$^2$ .

With Eqs. (\ref{eq:difco}), (\ref{eq:difco1}), and (\ref{eq:strvel}) in mind, the continuity equation takes the form,
\begin{equation}\label{eq:cont1}\frac{v_T}{n\sigma}\frac{dn}{dz}+\sqrt{\frac{g}{n\sigma}}n=0.\end{equation}
Integrating the latter yields
\begin{equation}\label{eq:nrar}n=\left[ n_c^{-1/2}+(\sqrt{g\sigma}/2\sqrt{2}v_T)(z-z_c)\right]^{-2}.\end{equation}
We observe that the characteristic exponential dependence of the standard barometric equation does not exist for dilute gases. In the deep rarified region of $z\gg z_c$, the dependence in Eq. (\ref{eq:nrar}) reduces to the form,
\begin{equation}\label{eq:deeprar}n=\frac{8k_BT}{mg\sigma}\frac{1}{z^2}.\end{equation}
Eq. (\ref{eq:deeprar}) predicts $n\sim n_c$ as should be expected from the adopted algorithm of matching the solutions in the barometric equation and the rarified gas regions.

Note that the coordinate dependencies in the barometric and rarified gas domains are described by two different sets of parameters, $(n_0,T,m,g)$ and $(\sigma , T,m,g)$ respectively. Therefore, our theory allows a possibility  of a relatively drastic change in $n(z)$ in the region of $z\sim z_c$ (although with the above numerical parameters in mind such a drastic change is not present).
\begin{figure}[t!]
\includegraphics[width=0.45\textwidth]{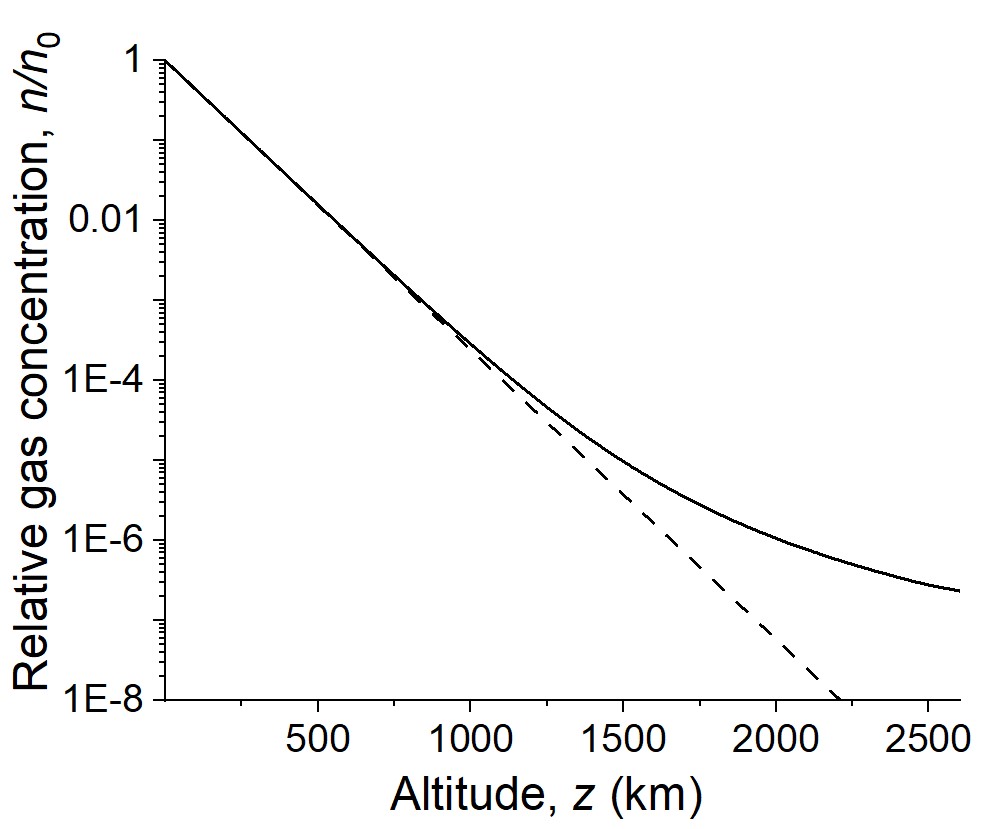}
\caption{Predicted distributions of the atmosphere density according to arguments of this section. The solid curve presents Eq. (\ref{eq:inteq}) with the following numerical parameters: $k_BT/mg=120$ km, $\sqrt{mg/k_BT\sigma n_0}=0.003$. The dashed line represents the barometric formula approximation.
\label{Fig:corona1}}
\end{figure}

Apart of the above mentioned drastic change, a semiquantitative formula interpolating between the limits of $z\ll z_c$ and $z\gg z_c$ regions can be obtained by interpolating
$v=v_Tv_g(v_T+v_g)^{-1}$, substituting which
in Eq. (\ref{eq:strvel}) yields,
\begin{equation}\label{eq:inteq}\ln\left(\frac{n}{n_0}\right)-2\sqrt{\frac{mg}{k_BT\sigma}}\left(\frac{1}{\sqrt{n}}-\frac{1}{\sqrt{n_0}}\right)=-\frac{zmg}{k_BT}.\end{equation}
Its predicted dependence is illustrated in Fig. \ref{Fig:corona1}. We observe the extended atmosphere tail in the rarified gas domain extending far beyond the barometric formula predictions.

\section{Long range behavior}\label{sec:LRB}
We now consider some predictions related to the gravity being a function of coordinates for the case of a spherically symmetric celestial body.  Taking into account the gravity universal law, the above equations remain applicable with the renormalization
\begin{equation}\label{eq:Newton}g\rightarrow g\left(\frac{R_0}{R}\right)^2\end{equation}
where $R_0$ is the body radius, and $R=R_0+z$ is the distance to its center. With the renormalization of Eq. (\ref{eq:Newton}) we obtain from Eq.(\ref{eq:cont1}),
\begin{equation}\label{eq:modif}\sqrt{\frac{v_T^2}{\sigma g}}\frac{dn}{n^{3/2}}=-\frac{R_0dR}{R} \quad {\rm when}\quad R\gg R_0 \end{equation}
where $g$ is the standard acceleration due to gravity at the celestial body surface. Eq. (\ref{eq:modif}) is readily integrated to yield
\begin{equation}\label{eq:atmdist}n=n_0\left[1+\Lambda\ln(R/R_0)\right]^{-2} \quad {\rm when}\quad R\gg R_0. \end{equation}
Here we have introduced,
\begin{equation}\label{eq:lambda}\Lambda = \sqrt{\frac{n_0\sigma mgR_0^2}{k_BT}}\sim 10^{4}-10^5.\end{equation}
Again, $g$ remains the acceleration due to gravity at the planet surface, and we assumed the following Earth related numerical values: $n_0\sim 10^{14}$ 1/cm$^3$, $T\sim 300 - 3000$ K,\cite{baliukin2019} $\sigma \sim 10^{-16}$cm$^2$,\cite{ruzic1985} $R_0=6000$ km, $m=3\times 10^{-24}$ g.

\begin{figure}[bh]
\includegraphics[width=0.47\textwidth]{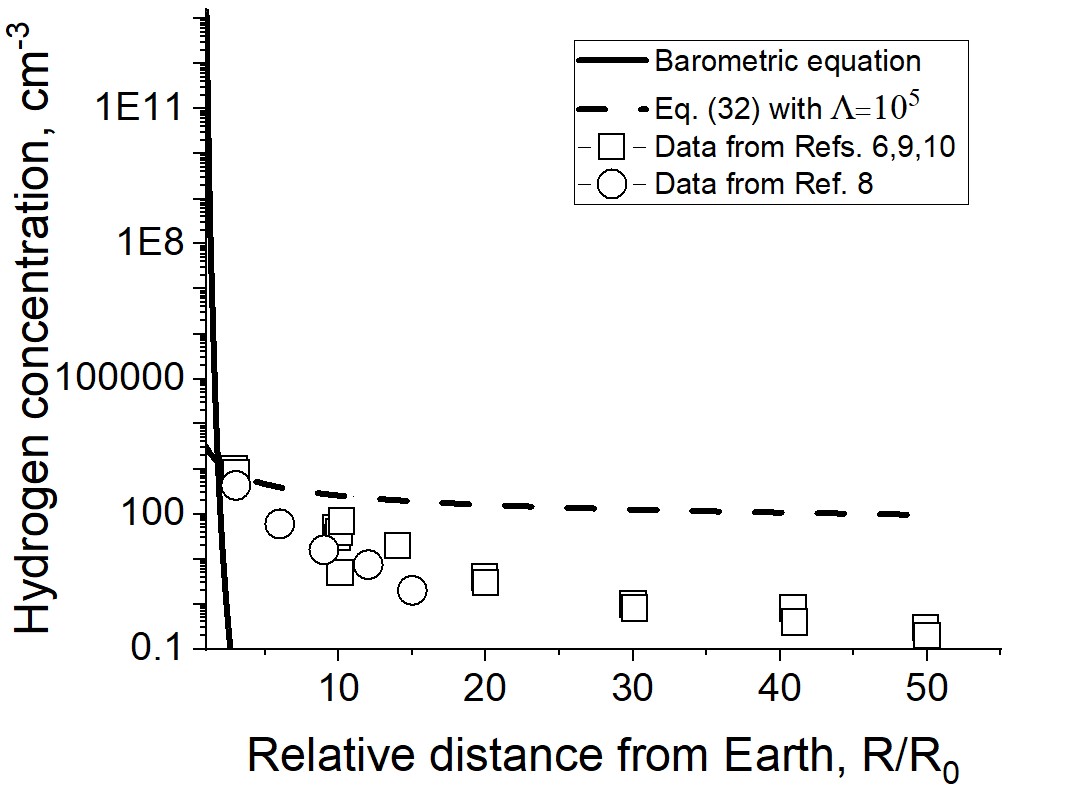}
\caption{The hydrogen density vs. $R/R_0$ from the barometric equation, Eq. (\ref{eq:atmdist}), and data from \cite{wallace1970, baliukin2019, padin2022,zoennchen2022}.
\label{Fig:atm_dens}}
\end{figure}

In Fig. \ref{Fig:atm_dens} we have plotted atmospheric density vs $R/R_0$ where the full line is the barometric equation, the dashed line is Eq. (\ref{eq:atmdist}) with  $\Lambda =10^5$, and the symbols are values extracted from the published data \cite{wallace1970, baliukin2019, padin2022,zoennchen2022}.  The most notable feature in In Fig. \ref{Fig:atm_dens} is that the dramatic change in slope in the density profile is consistent with experimental data. The latter reflects a qualitatively new situation that emerges when $R_0\gg L$: the characteristic value of $n$ for $R\gg R_0$ is much lower than $n_c$. We note the dramatic change in slope of the hydrogen density profile in Sun's corona  heuristically proposed in Ref. \cite{sakurai2017}.

%
%

The observed behavior of Terrestrial Exospheric profiles at several Earth radii ($R_E$) from the Earth center \cite{wallace1970, baliukin2019, padin2022,zoennchen2022} can be compared with the predictions of Eq. (\ref{eq:atmdist}). The existing geocoronal data derive the hydrogen concentration $n$, \cite{wallace1970,baliukin2019, padin2022,zoennchen2022} from optical measurements of the Lyman hydrogen emission assuming that the underlying Lyman excitations are due to Sun light. It is customary to describe the data with an empirically established dependence $R^{-3}$ whose physical explanation was attempted in Ref. \cite{baliukin2019} as described in Appendix \ref{sec:boltz}.

Our interpretation here relates the geocorona radiation to the gravitational energy of the falling hydrogen transferred to the atmosphere in the range of heights around $z_c$ (see Table \ref{tab:par}) where interatomic collisions dominate. As estimated in Eq. (\ref{eq:power}), the power so released, $w\sim 0.1$ $\mu$W/cm$^2$ turns out to be two-tree orders of magnitude higher than the measured. \cite{wallace1970, shklovsky1959} That discrepancy can be attributed to the average kinetic energy of a falling hydrogen atom (estimated as $\langle mv_g^2\rangle\sim {\cal P}/n_c\sim 0.1 $ eV) being insufficient to excite the Lyman series, so only rare atoms with energies much higher than the average can contribute to the geocorona glow. The same energy argument can possibly explain how the distribution of light emitting hydrogen in Fig. \ref{Fig:atm_dens} might have a somewhat different coordinate dependence compared to the total hydrogen concentration.

Incorporating the renormalization of Eq. (\ref{eq:Newton}) does not change the equations (\ref{eq:strvel}) and (\ref{eq:momflux}), so the momentum flux ${\cal P}=mnv_g^2$ remains coordinate independent. However, Eq. (\ref{eq:power}) acquires a renormalizing factor $R_0/R$ originating from the additional multiplier of $v_g$. For example, the energy flux into the Sun's atmosphere carried by the accreting hydrogen can be estimated as
\begin{equation}\label{eq:power1}w=\frac{mg_S}{4\sigma}\sqrt{\frac{g_SL_S}{2}}\frac{R_0}{R}.\end{equation}
Here, any quantity with a subindex $S$ indicates a parameter of the Sun. Using the gravity law, it is straightforward to see that $g_S\approx 30\times g$ and $L_S\sim L$. Therefore the Sun related energy flux is about 100 times higher than that of Earth, $w_{\rm Sun}\sim 10$ $\mu$W/cm$^2$ at its surface.

With the above in mind, extending the above theory to the case of Sun's corona, results in the prediction $w\sim 1-10 $ $\mu$Wcm$^{-2}$, in fair agreement with the data. \cite{leinert1998,kimura1998,pinter2003} We attribute this better agreement to the fact that the solar acceleration due to gravity is about 30 times greater than that of Earth. Therefore the energy of falling hydrogen atoms may be sufficient to excite the Lyman series for the case of Sun.

\section{Caveats and other possible applications}\label{sec:app}

An implicit simplification in our treatment is that the temperature is determined externally.  This assumption ignores the heat produced by the release of gravitational potential energy of a falling gas even though its related energy flux was estimated in the above as $w\sim 0.1$ $\mu$W/cm$^2$ for Earth and $w\sim 10$ $\mu$W/cm$^2$ for Sun. Conceptually one could include the effects of this heating by noting that in a steady state solution the heat gain due to a condensing gas must be equal to the heat loss from the gas due to thermal conductivity and radiation. That formidable problem remains to be addressed.

Some insight can be obtained by noting that the increase in temperature caused by the accreting H$_2$ will be small close to the surface ($z \ll z_c$) because the incoming energy must be distributed to all the particles in the dense region.  Also the heat flux at large distances ($z \gg z_c$)  where $g \rightarrow 0$ will also lead to negligible heating. Therefore, there will be a region of max temperature at some altitude $z\sim z_c$.  Although qualitative, one can wonder if this hot zone might explain the existence of celestial coronas. Note that our proposed gravity source of energy can be masked but not eliminated by other processes, such as ionization, solar wind, etc.).

Another consequence of the instability we have found is that celestial bodies will be in a state of hydrogen accretion. This conclusion contradicts the usual belief that celestial bodies are all in the process of loosing their hydrogen atmospheres. In particular, our conclusion predicts that the gas giants that have grown to their present size will continue to grow until they reach a critical mass needed for nuclear ignition.
This concept of hydrogen accretion is also consistent with the repetitive NOVA explosions, such as a NOVA T Coronae Borealis, that explodes every $\approx 80$ years.\cite{NYT,NASA} (The period between explosions is shorter when a white dwarf is a component of binary system due to supply of hydrogen from its partner.)

\section{Conclusions}\label{sec:concl}

In summary, we have demonstrated the following:\\
1. The barometric equation has a finite range of applicability limited to the altitudes where the mean free path of molecules $l$ becomes comparable to the characteristic dimension $L$ determining the barometric predicted atmosphere decay. \\
2. In the complementary region, the atmosphere distribution is better described by the dilute gas laws, and its density exhibits power decay or even logarithmic decay vs. distance.\\
3. The atmosphere distribution exhibits an instability that corresponds to H$_2$ accretion that flows from great distances towards all stellar objects.\\
4. The accretion corresponds to a certain momentum flux and energy flux ($w\sim 0.1$ $\mu$W/cm$^2$ for Earth and $w\sim 10$ $\mu$W/cm$^2$ for Sun) that may contribute to and/or explain celestial coronas.\\
5. Other possible correlations include the observed long tails in the Earth's and Sun's atmospheres, the existence of gas giants that are still growing, and repetitive NOVA explosions.

To avoid any misunderstanding, the process of hydrogen accretion considered here does not rule out the known competing processes of hydrogen evaporation. The two trends must be carefully compared for each set of parameters and correlated with observations.

\section{Open Research}\label{sec:opres}
{\bf Data Availability Statement:} Data were not used, nor created for this theoretical research. Some published data presented for comparison with our results are found in Refs. \cite{baliukin2019,wallace1970,padin2022,zoennchen2022}.

*\section{Acknowledgements}We would like to thank Dr. Ken Gray and Professor Adolf Witt for enthralling discussions.
\bigskip

\appendix

\section{The stream velocity}\label{sec:streamvel}
Here, we consider the conditions under which the drift velocity is given by either the standard viscous form, $v_d=\mu F$ or the alternative $v_d=\sqrt{al}$ where $\mu$ is the mobility, $F$ is the force, $a=F/m$ is the acceleration, and $l$ is the mean free path. We build on a simple 1D treatment \cite{kittel} of the Brownian movement described by the equation of motion
\begin{equation}\label{eq:stoch}md^2x/dt^2=-\beta m(dx/dt) +mA(t)\end{equation}
where $\beta m(dx/dt)$ represents the viscous drag, and $mA(t)$ is a stochastic force.

The original treatment \cite{kittel} assumes totally stochastic force with zero average $\langle A\rangle =0$. Here, we consider the case of such a stochastic force superimposed on the non-vanishing component, i. e.
\begin{equation}A=A'+\overline{A} \quad {\rm with}\quad \langle A'\rangle =0, \quad \langle \overline{A}\rangle =a\delta (x-nl)\end{equation}
where $n=1,2,..$ is a natural number. The delta function in the latter equation models intermolecular collisions affecting the dynamic with the periodicity of mean free path $l$.

Following the original approach \cite{kittel} we then multiply Eq. (\ref{eq:stoch}) by $x$, substitute $x(d^2 x/dt^2)=(1/2)d^2(x^2)/dt^2 -(dx/dt)^2$. The subsequent averaging is performed by setting
\begin{equation} \label{eq:meanav}x=\overline{x}+\delta x\quad {\rm and} \quad v=dx/dt\end{equation}
with $\overline{x}$ representing the stream velocity and $\delta x$ standing for the superimposed stochastic movements, $\langle \delta x\rangle=0$. We neglect the viscous drag in the equation for stream velocity. In addition, because the diffusion and stream dynamics are mutually independent, we treat them separately. For the stream dynamics ($\overline{x}$) one gets,
\begin{equation}\label{eq:strdyn}\frac{1}{2}\frac{d^2\overline{x}^2}{dt^2}-\left(\frac{d{\overline x}}{dt}\right)^2-al=0.\end{equation}
Integrating the latter yields
\begin{equation}\label{eq:strveloc}\overline{x}=(v+al)t.\end{equation}
Here the constant of integration $v$ must be set to zero to satisfy the condition that $\overline{x}=0$ under zero acceleration. Thereby, we arrive at the ansatz of Eq. (\ref{eq:strvel}), to within the accuracy of insignificant numerical multiplier.

The equation for $\delta x$ remains the same as in the original treatment \cite{kittel}, in particular, $\langle \left( d\delta x/dt\right)^2\rangle =k_BT/m$ [Eq. (31.10) in Ref. \cite{kittel}].
The total velocity is then estimated as
\begin{equation}\label{eq:veltot} v_{\rm tot}=\sqrt{\frac{k_BT}{m}+gl}.\end{equation}
Remarkably, the same equation was heuristically proposed earlier \cite{grimsditch} based on the energy argument: $mv_{\rm tot}^2/2=k_BT/m+gl$.

\section{Atmospheric density through the Boltzmann equation}\label{sec:boltz}

The stationary kinetic Boltzmann equation [cf. Eq. (\ref{eq:Boltz})] was used to numerically model the Earth's atmosphere density \cite{baliukin2019} for the following three models: Model 1 neglecting both the collision and ionization processes, Model 2 accounting for ionization but not for collisions, and Model 3 adding the effects of radiation pressure. The results of such modeling is that Model 1 approximately reproduces the often observed $R^{-3}$ dependence. That presents the first explanation for the data obtained from optical observations. 

In view of its importance, we briefly revisit the approach \cite{baliukin2019} here. We limit ourselves to the Model 1, in which case the stationary Boltzmann equation simplifies to the form,
\begin{equation}\label{eq:Boltz2}
v\frac{\partial f}{\partial R}-F\frac{\partial f}{\partial v}=0 \quad {\rm with}\quad F=-mg\frac{R_0^2}{R^2},\end{equation}
allowing separation of variables, $f=n(R){\cal V}(v)$. Substituting the latter in Eq. (\ref{eq:Boltz2}) yields,
\begin{equation}\label{eq:Boltz3} \frac{1}{Fn}\frac{\partial n}{\partial R}=\frac{1}{v{\cal V}}\frac{\partial {\cal V}}{\partial v}=const.\end{equation}
The constant in Eq. (\ref{eq:Boltz3}) is determined from the boundary condition ${\cal V}\propto \exp (-mv^2/k_BT)$ at the planet surface. As a result, $n$ is given by Eq. (\ref{eq:bareq2}). It is our understanding that the latter is plotted in Fig. 5 of Ref. \cite{baliukin2019} in comparison with $r^{-3}$.

To identify the slop of $n(R)$ from Eq. (\ref{eq:bareq2}) with that of $R^{-3}$ their logarithmic derivatives must coincide, which yields,
\begin{equation}\label{eq:logd}3LR=R_0^2.\end{equation}
Using numerical values $L=120$ km and $R_0=6000$ km Eq. (\ref{eq:logd}) yields $R\approx 15R_0$. This value is within the range of measured distances, and thus the two slopes can look visually similar in that range being quite different in general.

\end{document}